\documentclass[12pt]{article}
\pdfoutput=1
\usepackage{color,amsmath,amssymb,exscale,psfrag,epsfig}
\usepackage{cite,color,url}
\usepackage{pdfpages}	
\usepackage{graphicx}
\usepackage{slashed}
\usepackage[utf8x]{inputenc}
\usepackage[T1]{fontenc}
\usepackage{xcolor}
\usepackage{accents}
\usepackage{centernot}
\usepackage{fixltx2e}
\usepackage{enumerate}
\usepackage{float}
\usepackage{subfig}
\usepackage{floatpag}
\usepackage{placeins}
\floatpagestyle{empty}

\newcommand{\gel}{g_{el}}
\newcommand{\gc}{g_*}

\usepackage[colorlinks=true
,urlcolor=blue
,anchorcolor=blue
,citecolor=blue
,filecolor=blue
,linkcolor=blue
,menucolor=blue
 ,pagecolor=blue
,linktocpage=true
,pdfproducer=medialab
,pdfa=true
]{hyperref}
\def\bea{\begin{eqnarray}}
\def\eea{\end{eqnarray}}

\definecolor{nicered}{rgb}{0.7,0.1,0.1}
\definecolor{nicegreen}{rgb}{0.1,0.5,0.1}

\newcommand{\zorro}{$Z'$-explorer~}
\newcommand{\zorrobis}{$Z'$-explorer}

\catcode`\@=11
\def\lsim{\mathrel{\mathpalette\@versim<}}
\def\gsim{\mathrel{\mathpalette\@versim>}}
\def\@versim#1#2{\vcenter{\offinterlineskip
\ialign{$\m@th#1\hfil##\hfil$\crcr#2\crcr\sim\crcr } }}
\catcode`\@=12
\catcode`@=12
\DeclareUnicodeCharacter{2212}{-}

\begin{document}
\thispagestyle{empty}
\begin{flushright}
ICAS 050/20
\end{flushright}
\vspace{0.1in}
\begin{center}
{\Large \bf \zorro: a simple tool to probe $Z'$ models\\ against LHC data} \\
\vspace{0.2in}
	{\bf Ezequiel Alvarez$^{\dagger}$,
	Mariel Estévez$^{\diamond}$,
	Rosa Mar\'ia Sand\'a Seoane$^{\star}$,
}
\vspace{0.2in} \\
	{\sl International Center for Advanced Studies (ICAS)\\
 UNSAM, Campus Miguelete, 25 de Mayo y Francia, (1650) Buenos Aires, Argentina }
\\[1ex]

\end{center}
\vspace{0.1in}

\begin{abstract}
New Physics model building requires a vast number of cross-checks against available experimental results.  In particular, new neutral, colorless, spin-1 bosons $Z'$, can be found in many models. We introduce in this work a new easy-to-use software {\it \zorro} which probes $Z'$ models to all available decay channels at LHC.  This program scrutinizes the parameter space of the model to determine which part is still allowed, which is to be shortly explored, and which channel is the most sensitive in each region of parameter space.   User does not need to implement the model nor run any Monte Carlo simulation, but instead just needs to use the $Z'$ mass and its couplings to Standard Model particles.  We describe \zorro backend and provide instructions to use it from its frontend, while applying it to a variety of $Z'$ models.  In particular we show \zorro application and utility in a sequential Standard Model, a B-L $Z'$ and a simplified two-sector or Warped/Composite model.  The output of the program condenses the phenomenology of the model features, the experimental techniques and the search strategies in each channel in an enriching outcome.  We find that compelling add-ons to the software would be to include correlation between decay channels, low-energy physics results, and Dark Matter searches.  The software is open-source ready to use, and available for modifications, improvements and updates by the community.
\end{abstract}

\vspace*{2mm}
\noindent {\footnotesize E-mail:
{\tt 
$\dagger$ sequi@unsam.edu.ar,
$\diamond$ mestevez@unsam.edu.ar.
$\star$ rsanda@unsam.edu.ar,
}
}

\newpage
\section{Introduction}
\label{section:1}

The Large Hadron Collider (LHC) is an outstanding and successful machine with countless achievements including the discovery of the Higgs boson \cite{Aad:2012tfa,Chatrchyan:2012xdj} as one of the most emblematic. In its current stage, the LHC is exploring unsurveyed energies and luminosities in its quest for discovering New Physics (NP).  One of the few conclusions that could be withdrawn from current results is that many theoretical forecasts have not been fulfilled so far, and searches should address a more phenomenological and perhaps general point of view.  

One expected NP signature in many NP models and also from a general phenomenological point of view is a neutral colorless spin-1 particle, also known as a $Z'$ \cite{Langacker:2008yv,Leike:1998wr} because of its similarity in quantum numbers to the Standard Model (SM) $Z$ boson.  The objective of this work is to develop a tool which can be used to test against current experimental results any model predicting a $Z'$ particle.  One of the main objectives pursued in this work is to create a very simple tool which could be learned and used rapidly to explore parameter space while constructing NP models or exploring pure $Z'$ phenomenology.  The only input required to test a $Z'$ model from this phenomenological point of view would be its couplings and/or branching ratios to the allowed decaying channels.

Because of the $Z'$ quantum numbers, its possible decay channels explored at the LHC and their corresponding latest results are  $jj$ \cite{Aad:2019hjw, CMS:2019oju, Aaboud:2018fzt}, $b\bar{b}$ \cite{Aaboud:2018tqo}, $t\bar{t}$ \cite{Sirunyan:2018ryr, Aaboud:2018mjh}, $e^{+}e^{-}$ \cite{CMS-PAS-EXO-19-019, Aad:2019fac}, $\mu^{+}\mu^{-}$ \cite{CMS-PAS-EXO-19-019, Aad:2019fac}, $\tau^{+}\tau^{-}$ \cite{Aaboud:2017sjh, Khachatryan:2016qkc}, $W^{+}W^{-}$ \cite{Aaboud:2017fgj, Sirunyan:2019vgt}, $Zh$ \cite{Sirunyan:2019vgt}.  Since testing all these channels while constructing a model it is cumbersome, we study in this manuscript a new tool which would facilitate this enterprise.  

To tackle this problem we observe that each decay channel has different sensitivity not only because of the specific detector and experimental techniques to reconstruct each particle, but also because of the different SM backgrounds affecting each final state, and because all of these depend on the mass of the sought $Z'$.  Although all these features for each channel are individually complex, and still more when combined, their relevant features are summarized in the exclusion plots presented by the experimental groups in each search.  We show that using this information we can estimate in a reasonable way whether a point in parameter space in a $Z'$ NP model is excluded or not, and which of the above is the most sensitive channel.  On this basis,  along this work we present the software {\it \zorro} which can make the mentioned computation automatically in a very simple manner and for a huge amount of points in parameter space.

There are available software which also aim to test NP models against available experimental data in a simple manner with connection to our proposal \cite{Kraml:2013mwa,Papucci:2014rja,Drees:2013wra,Dercks:2016npn,Barducci:2016pcb,Butterworth:2016sqg,Kahlhoefer:2019vhz}. Many of them are for specific NP models --such as for instance supersymmetry-- and others for specific processes --such as Dark Matter portal.  Refs.~\cite{Barducci:2016pcb} and \cite{Kahlhoefer:2019vhz} are designed for general purpose $Z'$ models.  Ref.~\cite{Barducci:2016pcb} is a very complete software analyzing a variety of NP signatures, however the $Z'$ module only addresses the dilepton and dijet channels.  On the other hand, recent Ref.~\cite{Kahlhoefer:2019vhz} also includes interesting finite width and interference effects, however its current analysis is restricted to dilepton final states.  In the present manuscript, although our mechanism is more simplified than others in some aspects, it is reasonably solid and we analyze simultaneously all possible $Z'$ decay channels while providing the quantitative strength for each one of them.

This article is organized as follows. In Sect.~\ref{section:2} we briefly review LHC $Z'$ searches and their sensitivity, also focusing on some other specific topics relevant for our purposes. In Sect.~\ref{section:3} we make a full description of our software. In Sect.~\ref{section:4}, we apply our results to three different models: the Sequential Standard Model (SSM), a B-L model and a two-sector or Warped/Composite Model. In Sect.~\ref{section:5} we present the outlook of our results, and discuss future possible prospects. Finally, Sect.~\ref{section:6} contains the conclusions.  We include in Appendix \ref{higgs}, as a working example, a computation of the strength for the Higgs boson discovery using the previous year data and predict which channel would discover it.  

\section{LHC Z' searches and sensitivity}
\label{section:2}

The aim of this article is to present \zorro, a simple software that determines the most sensitive channel for the exclusion (or detection) of a $Z'$ boson.  \zorro can analyze any point in parameter space of NP models given by the user.  With the purpose of better understanding \zorro functioning as explained in next sections, we briefly describe in the following paragraphs how LHC $Z'$ searches are addressed and introduce related topics needed to understand the software building.

A $Z'$ resonance is produced at the LHC through quark-antiquark annihilation.  Therefore its production cross-section depends on its coupling to valence and sea quarks, $u$, $d$, $c$, $s$ and $b$, which depend on the point in parameter space in the NP Lagrangian.  Since $Z'$ is a spin-1 particle, it cannot be produced at loop level through gluon fusion because of Landau-Yang theorem~\cite{Landau:1948kw,Yang:1950rg,Fox:2018ldq}

$Z'$ decay modes and branching ratios also depend on the specific NP model.  $Z'$ decay to dijets depends on the above mentioned coupling to quarks, but its branching ratios also depends on the $Z'$ coupling to all other particles
The search of a $Z'$ at the LHC, as for any other NP particle, is implemented in different decay channels. At the LHC, searches have been performed both by the CMS and the ATLAS Collaborations, and have superseded the results coming from searches in other colliders such as Tevatron in most of the TeV scale \cite{Tanabashi:2018oca}. 

Although the fraction of times a $Z'$ decays to a given channel depends solely on its corresponding branching ratio, in experimental terms the acceptance and efficiency of the detector and the background of this channel are crucial to determine its exclusion/discovery sensitivity. These features also have an important dependence on the scale of energies considered, which depends on the sought resonance mass. For instance, for the search of a $Z'$ with $M_{Z'}\lesssim 1$ TeV at the LHC, we expect the dijet channel to have large QCD background, which dominates at low $p_{T}$. Whereas in the dilepton channel, the background is smaller in comparison, and thus the signal is likely cleaner and the channel more sensitive even though its branching ratio could be smaller.    Considering also the $W^{+}W^{-}$ decay mode, for the same above reasons the fully hadronic decay is hidden in QCD background, whereas the leptonic decays, although cleaner, are difficult to reconstruct due to missing energy.  This scenario changes for larger masses, $M_{Z'}\gtrsim 2$ TeV.  In such a case there is a more subtle interplay between the features discussed above.  Not only because all backgrounds are highly reduced, but also because the techniques for detecting and reconstructing particles are also modified.  In particular, in this high $p_{T}$ regime, boosted massive particles, such as $W$ boson or $t$ quark, can decay hadronically in configurations where all particles are collimated in a single fat jet, which is more distinguishable from QCD background.  This, in conjunction to a reduced QCD background yields a favorable enhancement for channels such as hadronic $W$ and top.   

From this overview on some of the features involved in determining which is the best channel to exclude (or find) a given $Z'$, we see that there are a variety of non-trivial ingredients which affect which channel could be the most sensitive.  It is of particular interest for our work to determine a magnitude that condenses all the relevant information and provides an estimation on the sensitivity of each channel for each point in parameter space in a given NP $Z'$ model. From the phenomenological point of view, the information available on previous experimental searches in the different channels at a given energy and luminosity can be incorporated into the determination of the most sensitive decay channel, through the extraction of the predicted experimental limits. With this, the strength ($\mathcal{S}$) of the signal of each channel can be defined as \cite{Alvarez:2016ljl}
\begin{equation}
\mathcal{S} = \frac{\sigma_{pred}}{\sigma_{lim}}
\label{s}
\end{equation}
where $\sigma_{pred}$ is the predicted cross-section times branching ratio times acceptance of the new $Z'$, and $\sigma_{lim}$ is the corresponding predicted experimental upper limit at the $95\%$ CL. Despite the complex compromise between the expected theoretical branching ratio and the cleanliness of the experimental signal coded in its definition, the strength {\cal S} has a simple interpretation. If for a given channel $\mathcal{S} >1$ for a given point in the parameter space, the point is experimentally excluded. But if $\mathcal{S}<1$ for all channels, the point is not only not excluded but also the one with the largest $\mathcal{S}$ is expected to be the most sensitive channel for the exclusion or observation of the particle being sought.  We illustrate this statement in Appendix \ref{higgs} by using the data from the year previous to the Higgs discovery.

The difference in using {\it predicted} instead of {\it measured} limits consists in considering the experimental techniques instead of eventual fluctuations in the data.  In case the  measured limit departs significantly from the predicted value, it would correspond to a discovery or a wrong prediction which in both cases should be addresses from another point of view.

Ideally, for this comparative analysis between all possible final states for a particle, it is expected that for a given energy the experimental measurements for each channel will be at the same luminosity, but this is not always the case. However, these luminosities tend to be almost within the same order of magnitude, so assuming a statistical uncertainty regime, the experimental sensitivity can be rescaled with the square root of the ratio of the luminosities \cite{Alvarez:2016ljl}. 
\begin{equation}
\sigma_{lim}^{(2)}=\frac{\sigma_{lim}^{(1)}}{\sqrt{L_{2}/L_{1}}}
	\label{luminosity}
\end{equation}
See also Appendix \ref{higgs} where this assumption is verified in a real case scenario.


\section{\zorro software}
\label{section:3}
Motivated by the discussion and strength ($\mathcal S$) definition in previous section, we have designed a simple software that, for a given $Z'$ NP model, it can compare the sensitivity of available channels.  Along this section we describe how this software works and how it can be utilized by any user to extract useful and relevant information for any $Z'$ model.

\subsection{Backend}
\label{bend}

\zorro is a {\tt C++} open source program \cite{github} that allows to determine the most sensitive channel for the search of a $Z'$ boson at LHC as a function of the relevant NP parameters in the model: the mass of the $Z'$ boson ($M_{Z'}$), the couplings of $Z'$ to all SM-fermions and the decay widths to diboson and neutrinos and/or invisible.   

\zorro running flow is quite simple.  Using the above input parameters, the program computes the production cross section for $Z'$ and the different decay rates to each one of the considered channels, which are: $jj$, $b\bar{b}$, $t\bar{t}$, $e^{+}e^{-}$, $\mu^{+}\mu^{-}$, $\tau^{+}\tau^{-}$, $\nu\bar\nu$ (or any invisible), $W^{+}W^{-}$ and $Zh$. Once this is obtained, these predictions for each possible final state are compared to the corresponding $95\%$ upper limit on the product of the cross section times branching fraction, coming from the most sensitive searches performed both by the ATLAS and CMS collaborations. The comparison is made through the calculus of the aforementioned strength $\mathcal{S}$, explained in Sect.~\ref{section:2}.

To compute the production cross section, the program uses previously generated and recorded production cross section with {\tt MadGraph5\_aMC@NLO} \cite{Alwall:2014hca} with a tailored $Z'$ model which couples with unity to only one quark in the proton each time and for a set of values of $M_{Z'}$ between 0.4 and 8.1 TeV, with a step of 0.01 TeV, and at $\sqrt{s}=13$ TeV.   Since LHC protons are unpolarized we set same couplings for Left and Right chiralities at this level.   These calculus, stored in the repository as {\tt /cards/sim\_cards}, are used during program execution: the predicted production cross section for a specific reference point is the sum of the five contributions of quarks ($u$, $d$, $c$, $s$ and $b$).  Each of them adjusted by the sum of the corresponding squared chiral couplings, that are extracted from the input card. The software selects inside the simulations the record with the mass $M_{Z'}$ that is closest to the one in the input card at the corresponding reference point. The total production cross section is then used in each possible decay channel to compute the $\sigma_{pred}$ needed in Eq.~\ref{s}, multiplying by the corresponding branching ratio, which is calculated by the program using the input parameters.

The experimental limits used by the program, corresponding to the value of $\sigma_{lim}$ required in Eq.~\ref{s}, are stored in {\tt /cards/exp\_cards}. The searches included in the program, performed by both ATLAS and CMS collaborations and considering the latest results up to date, include the following channels and references: $jj$ \cite{Aad:2019hjw}, $b\bar{b}$ \cite{Aaboud:2018tqo}, $t\bar{t}$ \cite{Sirunyan:2018ryr}, $e^{+}e^{-}$ \cite{CMS-PAS-EXO-19-019}, $\mu^{+}\mu^{-}$ \cite{CMS-PAS-EXO-19-019}, $\tau^{+}\tau^{-}$ \cite{Aaboud:2017sjh}, $W^{+}W^{-}$ \cite{Aaboud:2017fgj}, $Zh$ \cite{Sirunyan:2019vgt}. We use the digitalization \cite{Uwer:2007rs} of the numerical values of $95\%$ upper limits on the product of the cross section and the branching fraction versus $M_{Z'}$ from the corresponding published plots.  It is worth noticing that since searches are constantly updated, the user can modify the experimental limits for a given channel just by modifying the corresponding {\tt exp\_card} and propose the update of the repository.  The program includes a {\tt README} file in which the numeric labeling of {\tt exp\_card} for each channel is displayed.

\subsection{Frontend}
\label{fend}
One of the main features of our software is its simplicity. The user only has to provide an input card as {\tt /incard/card\_1.dat}, specifying $M_{Z'}$ (in TeV), the $Z'$ couplings to all SM-fermions (except for neutrinos, which information is required through $\Gamma_{inv}$), and the partial widths to $W^{+}W^{-}$ ($\Gamma_{WW}$) and $Zh$ ($\Gamma_{Zh}$).  We require this latter information as {\it partial width} since there is not a unique Lorentz structure in their couplings.  The total width to non SM particles can be added in the computation as $\Gamma_{xx}$.  The data should be specified as one row for each point in parameter space with the following order in each row:
\begin{equation}
M_{Z'} \hspace{0.1cm} g_{u_L} \hspace{0.1cm} g_{u_R} \hspace{0.1cm} g_{d_L} \hspace{0.1cm} g_{d_R}\hspace{0.1cm} g_{c_L} \hspace{0.1cm} g_{c_R} \hspace{0.1cm} g_{s_L} \hspace{0.1cm} g_{s_R}\hspace{0.1cm} g_{b_L} \hspace{0.1cm} g_{b_R} \hspace{0.1cm} g_{t_L} \hspace{0.1cm} g_{t_R}
\hspace{0.1cm} g_{e_L} \hspace{0.1cm} g_{e_R} \hspace{0.1cm} g_{\mu_L} \hspace{0.1cm} g_{\mu_R}\hspace{0.1cm} g_{\tau_L}\hspace{0.1cm}g_{\tau_R}\hspace{0.1cm}\Gamma_{inv}\hspace{0.1cm}\Gamma_{WW}\hspace{0.1cm}\Gamma_{Zh}\hspace{0.1cm}\Gamma_{xx}
\nonumber
\end{equation}
where $g_{f_L}$ ($g_{f_R}$) is the coupling of $Z'$ to the corresponding Left (Right) fermion. There is no limit in the number of rows in the input card, and the program runs very fast and using negligible CPU resources.  Each row, with its corresponding twenty-three columns, is a different reference point in the parameter space for the program.

To execute the program, once in the main directory in a Linux terminal, the user has to write in command line the following instruction:
\begin{equation}
./program.out
\nonumber
\end{equation}
A brief summary of the tasks performed by \zorro will be displayed on screen. The generated output file (saved in {\tt /output/1.dat}), contains the same information than the input card in each line, to facilitate the subsequent processing and analysis of the generated information, followed by the calculated strength ($\mathcal{S}$) in each channel, presented in the following order from column twenty-four to column thirty-three:
\begin{equation}
\mathcal{S}_{jj}\hspace{0.2cm}\mathcal{S}_{bb}\hspace{0.2cm}\mathcal{S}_{tt}\hspace{0.2cm}\mathcal{S}_{ee}\hspace{0.2cm}\mathcal{S}_{\mu\mu}\hspace{0.2cm}\mathcal{S}_{\tau\tau}\hspace{0.2cm}\mathcal{S}_{inv}\hspace{0.2cm}\mathcal{S}_{WW}\hspace{0.2cm}\mathcal{S}_{Zh}\hspace{0.2cm}\mathcal{S}_{xx}\hspace{0.2cm}
\nonumber
\end{equation}
Observe that the $\mathcal{S}_{xx}$ is left as a dummy variable to eventually add new possible $Z'$ channels in the future.

Since each row in the input file is a different point in the parameter space for \zorrobis, output card contains the same number of rows of data than {\tt card\_1.dat}. The running time depends on this number of rows and the CPU speed, but typically for an input card of $1000$ reference points, time is less than $2$ seconds.

Additionally, after running, in the {\tt /extra} folder and in different files are available the estimated decay widths, branching ratios, the estimated $\sigma_{pred}$ and the extracted $\sigma_{lim}$ for each reference point in the input card. More details on these auxiliary files can be found in the {\tt README} file.

\section{\zorro application examples}
\label{section:4}

Along this section we implement the software on a few well known $Z'$ models to show its potential and utility.  We do not aim to explore in detail NP models and place constraints, but instead to show the software scope through some concrete simplified examples.  As a matter of fact, we explore some regions in parameter space that are already discarded by other sectors, however we find it suitable for the sake of showing the richness \zorro would provide in that case.  We utilize \zorro on a Sequential SM, on a B-L model and on a Warped/composite scenario.

\subsection{Sequential Standard Model}
\label{ssm}
As a first example to test the power of \zorro software, we present an analysis for the Sequential Standard Model (SSM) \cite{Altarelli:1989ff}, where the couplings of the $Z'$ boson to all SM fermions are equal to those of the $Z$ boson. The SSM is a reference model in experimental searches for its simplicity, and we use it as validation for our program.

\begin{figure}[h!]
\begin{center}
\includegraphics[width=0.75\textwidth]{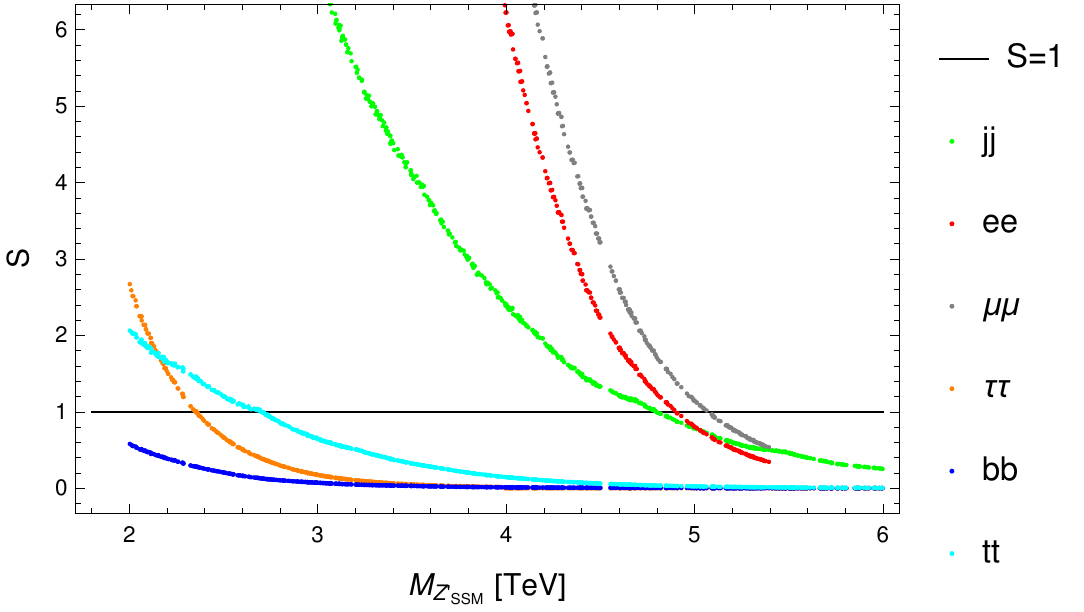}
\caption{\small Strength ($\mathcal{S}$) for each fermionic decay channel included in \zorro for a $Z'_{SSM}$. Experimental limits used by the program are the aforementioned in Sect.~\ref{section:3}. The line corresponding to $\mathcal{S}=1$ delimits the excluded region.}
\label{ssmplot}
\end{center}
\end{figure}

In Fig.~\ref{ssmplot} we present the strength $\mathcal{S}$ to all the fermion decay channels included in \zorro in terms of $M_{Z'}$. The comparison of the sensitivities between the different experimental analyses is straightforward.  Since couplings are not suppressed for light fermions, as it can be expected the dimuon and dielectron channels are the most sensitive.  In particular, these results match those obtained by CMS in Ref.~\cite{CMS-PAS-EXO-19-019}.   Comparing these two lepton channels, since experimental efficiency is slightly  better for muon than electrons, there is more sensitivity in the dimuon decay channel as expected.   It is interesting to notice how the $t\bar t$ channel passes in strength the $\tau\bar\tau$ channel for masses greater than $\sim 2$ TeV.   The reason behind this are the well known developed experimental techniques for boosted top tagging \cite{Aaboud:2018psm}.  Another interesting feature to be observed is how the dijet channel strength falls down slower than dilepton channels for larger $Z'$ masses, even surpassing the di-electron channel above 5 TeV.  This could be associated to the backgrounds: the dijet background is mostly related to the gluon PDF in the proton, whereas the dilepton background is to the quark content in the proton PDF, and is well known that valence quarks are more likely than gluons at large $x$.   This kind of features show some aspects of all the information condensed in the strength $\mathcal{S}$.

Additionally, we also carried out \zorro validation using the latest ATLAS results~\cite{Aad:2019fac}, where the information for both leptonic channels is in the combined dilepton channel.

\subsection{B-L model}
\label{b-l}

We present the sensitivity analysis for the search at LHC of a $Z'$ coming from a minimal $U(1)$ extension of the SM, associated to the B-L number. The gauge sector Lagrangian for this NP scenario is \cite{Gutierrez-Rodriguez:2015qka, Basso:2011hn, Khalil:2006yi}

\begin{equation}
{\cal L}_{g} = -\frac{1}{4} B_{\mu\nu}B^{\mu\nu}-\frac{1}{4} W_{\mu\nu}^{a}W^{a\mu\nu}-\frac{1}{4} Z'_{\mu\nu}Z'^{\mu\nu}
\end{equation}
where $W_{\mu\nu}^{a}$, $B_{\mu\nu}$ and $Z'_{\mu\nu}$ are the field strength tensors for $SU(2)_{L}$, $U(1)_{Y}$ and $U(1)_{B-L}$ (ignoring the $SU(3)_C$ gauge symmetry for the purposes of our analysis).

In this type of models, $Z-Z'$ mixing occurs through the conventional spontaneous symmetry breaking (SSB)
mechanism, because the usual Higgs doublet is not singlet under the new $U(1)_{B-L}$ \cite{Rizzo:2006nw}. The scalar sector is defined as
\begin{equation}
{\cal L}_{s} = (D^{\mu}\Phi)^\dagger(D_{\mu}\Phi)+(D_{\mu}\chi)^\dagger(D_{\mu}\chi)-V(\Phi,\chi)
\end{equation}
with the potential defined as \cite{Khalil:2006yi,Basso:2010jm}

\begin{equation}
V(\Phi,\chi) = m^2(\Phi^\dagger\Phi)+\mu^2|\chi|^2+\lambda_{1}(\Phi^\dagger\Phi)^2+\lambda_{2}|\chi|^4+\lambda_{3}(\Phi^\dagger\Phi)|\chi|^2
\label{potencial}
\end{equation}
where $\Phi$ is the aforementioned scalar doublet, and $\chi$ is a complex singlet scalar field required to achieved the breaking of the new gauge group to acquire a {\it vev} at the TeV scale. The covariant derivatives are \cite{Basso:2010jm, Gutierrez-Rodriguez:2015qka}
\begin{eqnarray}
D^{\mu}\Phi &=& \partial_{\mu} \Phi+i\left[\frac{g}{2} T^{a} W_{\mu}^{a}+ g_{1}Y B_{\mu}+ g_{1}' Y' B_{\mu}'\right]\Phi\nonumber \\
D^{\mu}\chi &=& \partial_{\mu} \chi+i g_{1}' Y' B_{\mu}'\chi
\end{eqnarray}
with $g=e/\sin(\theta_{W})$ and $g_{1}=e/\cos(\theta_{W})$. After SSB, the scalar fields can be written as
\begin{gather}
\Phi=
\left(\begin{array}{c}
0\\
\frac{v+\phi^{0}}{\sqrt{2}}\end{array} \right),
\hspace{0.5cm}
\chi=\frac{v'+\phi^{'0}}{\sqrt{2}}
\label{ssb}
\end{gather}
where the two {\it vev}'s, $v$ and $v'$, are real and positive. From the minimization of Eq.~\ref{potencial} one obtains the mass matrix for $\phi^{0}$ and $\phi^{'0}$, and after its diagonalization the mass eigenstates $h$ and $H$ are obtained
\begin{gather}
\left(\begin{array}{c}
h\\
H\end{array} \right)=
\left(\begin{array}{cc}
\cos (\alpha) & -\sin (\alpha)\\
\sin (\alpha) &  \cos (\alpha) \end{array} \right)
\left(\begin{array}{c}
\phi^{0}\\
\phi^{'0}\end{array} \right) .
\label{alpha}
\end{gather}
Here $\alpha$ is the mixing angle and $h$ is the SM-like Higgs boson. Along this work, we consider $\cos (\alpha)\simeq 1$ and thus $h\simeq\phi^{0}$ and $v=0.246\ TeV$. 

The interactions vertices between the neutral gauge bosons $Z$, $Z'$ and a pair of SM fermions are given by
\cite{Gutierrez-Rodriguez:2015qka}

\begin{equation}
{\cal L}_{NC} =-\frac{ig}{\cos(\theta_{W})}\sum_{f}\bar{f}\gamma^{\mu} \frac{1}{2}(g_{v}^{f}-g_{a}^{f}\gamma^5)f Z_{\mu}-\frac{ig}{\cos(\theta_{W})}\sum_{f}\bar{f}\gamma^{\mu} \frac{1}{2}(g_{v}^{'f}-g_{a}^{'f}\gamma^5)f Z_{\mu}'
\end{equation}
where the new couplings are
\begin{equation}
	\begin{aligned}
		g_{v}^{f} &= T_{3}^f \cos(\theta_{B-L})-2 Q_{f}\sin^2(\theta_{W})\cos(\theta_{B-L})+2 Y' \frac{g_{1}'}{g}\cos(\theta_{W})\sin(\theta_{B-L}) \\
g_{a}^{f} &= T_{3}^f \cos(\theta_{B-L})  \\
g_{v}^{'f} &= -T_{3}^f \sin(\theta_{B-L})+2 Q_{f}\sin^2(\theta_{W})\sin(\theta_{B-L})+2 Y' \frac{g_{1}'}{g}\cos(\theta_{W})\cos(\theta_{B-L}) \\
g_{a}^{'f} &= -T_{3}^f \sin(\theta_{B-L}) 
	\end{aligned}
	\label{couplingsbl}
\end{equation}
and $\theta_{B-L}$ is the $Z$-$Z'$ mixing angle.  In the limit where $g_{1}'=0$ and $\theta_{B-L}=0$, the couplings of the $Z$ boson to all SM fermions are recovered.  This angle is bounded by LEP searches, and the current bound on this parameter is $|\theta_{B-L}|\leq10^{-3}$ \cite{Tanabashi:2018oca}.  Electro-Weak Precision Tests from LEP-II also place limits in the model and therefore we restrict our exploration to the 99\% C.L.~allowed region $M_{Z'}/g_1 \geq 7.1$ TeV \cite{Cacciapaglia:2006pk}

The additional formulae required by \zorro are the decay widths of the new $Z'$ boson to $W^{+}W^{-}$ \cite{Leike:1998wr,PhysRevD.36.3429} and to $Zh$\cite{Barger:2009xg,PhysRevD.36.3429}
\begin{equation}
	\begin{aligned}
	\Gamma (Z'\rightarrow W^{+}W^{-}) & =  \frac{g^{2}}{192 \pi}\cos^2(\theta_{W})\sin(\theta_{B-L})M_{Z'} \left(\frac{M_{Z'}}{M_Z}\right)^4\left(1-4\frac{M_{W}^2}{M_{Z'}^2}\right)^{3/2}\\ 
	&\ \times \left[1+20\frac{M_{W}^2}{M_{Z'}^2}+12\frac{M_{W}^4}{M_{Z'}^4}\right]
	\end{aligned}
\end{equation}
\begin{equation}
	\begin{aligned}
		\Gamma (Z'\rightarrow Zh)&=\frac{g^{2}M_{Z}^2}{192 \pi M_{W}^2} M_{Z'}\sqrt{\lambda}\left(\lambda+12\frac{M_{Z}^2}{M_{Z'}^2}\right)\\ &\ \times \left[\left(\frac{4M_{Z}^2}{v^2}-g_{1}'^2\right)\sin(2\theta_{B-L})+\left(\frac{4M_{Z}g_{1}'}{v}\right)\cos(2\theta_{B-L})\right]
	\end{aligned}
\end{equation}
with
\begin{equation}
\lambda(1,\frac{M_Z^2}{M_{Z'}^2},\frac{M_h^2}{M_{Z'}^2})=1+\left(\frac{M_Z^2}{M_{Z'}^2}\right)^2+\left(\frac{M_h^2}{M_{Z'}^2}\right)^2-2\left(\frac{M_Z^2}{M_{Z'}^2}\right)-2\left(\frac{M_h^2}{M_{Z'}^2}\right)-2\left(\frac{M_Z^2}{M_{Z'}^2}\right)\left(\frac{M_h^2}{M_{Z'}^2}\right)
\end{equation}

\begin{figure}[h!]
\begin{center}
\includegraphics[width=0.75\textwidth]{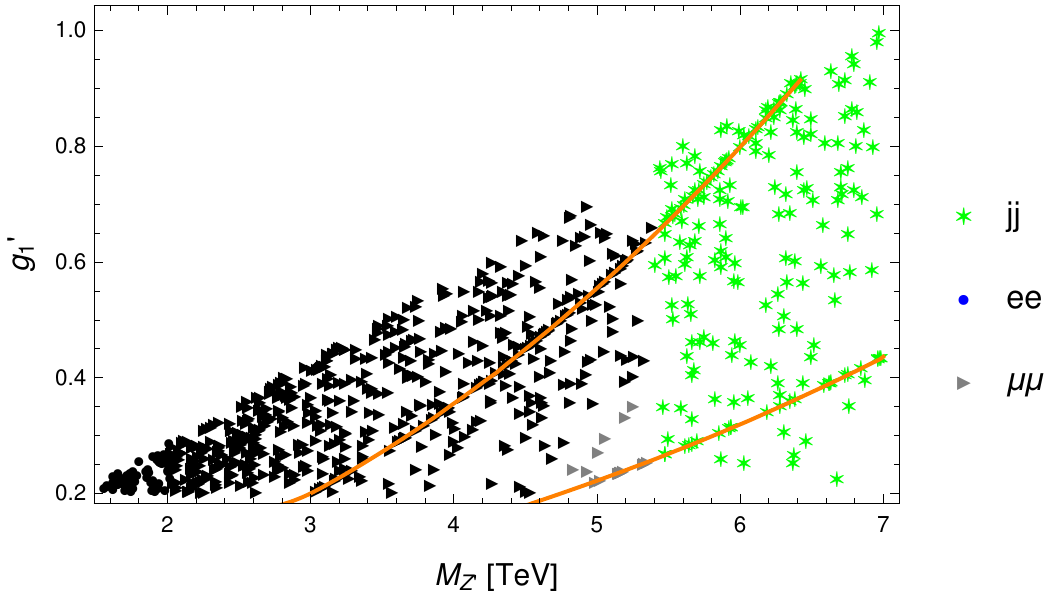}
\caption{\small Most sensitive channel in each region of the $(g^{'}_{1}, M_{Z'})$ parameter space for the search of the $Z'_{B-L}$ described in Sect.~\ref{b-l}. Experimental limits used by the program are the aforementioned in Sect.~\ref{section:3}. The points in black are experimentally excluded by the channel indicated by their shape. The orange lines correspond from left to right to the contour levels $\theta_{B-L}=2.5\times 10^{-4}$ and $10^{-4}$, respectively. }
\label{blplots}
\end{center}
\end{figure}

Using the above information we use \zorro to compute the strength for all channels as a function of $g_{1}'$ and $M_{Z'}$.  In Fig.~\ref{blplots} we present the most sensitive channel for each region of the parameter space. There is an area excluded mainly by dimuon searches for 2 TeV$\lesssim M_{Z'}\lesssim$ 5 TeV and 0.2$\lesssim g_{1}' \lesssim$ 0.7.  It is interesting to observe that the dielectron channel only excludes points with $M_{Z'}\lsim2$ TeV.  This is intimately related to the results in Ref.~\cite{CMS-PAS-EXO-19-019}, which is likely to be related to experimental details in the geometry and the detector's efficiencies.  In any case, the strength for dielectron and dimuon channels is very close in that region.   For small $g_{1}'$ and masses $\sim 5$ TeV there is a still not excluded region where dimuon is the most sensitive channel.  For large masses, where there are no experimental limits for leptonic final states, the dijet channel becomes the most sensitive channel. 


\subsection{Warped/Composite Model}
\label{warped}

As a third example we study the phenomenology of a $Z'$ coming from a warped/composite or two-sector model, through \zorrobis.  The goal in analyzing this model is to explore a NP scenario in which the $Z'$ couplings are related to the mass spectrum\footnote{A use of \zorro in a full Composite scenarios can be found in Ref.~\cite{lamagna}, which has been completed while this project was under development.}. We summarize a simplified version of a partial compositeness four dimensional model following Ref.~\cite{Contino:2006nn} lines of reasoning and notation, where more details can be found.

This model includes, beside the ones in the SM, new fermions and bosons with masses around the TeV scale.  The composite sector contains vector excitations $\rho _{\mu}$ which respect a $ SU(3)_C \otimes SU(2)_L$ $\otimes SU(2)_R \otimes U(1)_X$ symmetry, fermions $\chi$ and $\tilde\chi$ --one for each elementary fermion--, and the Higgs field $H$.  We use the same fermion embedding as in Ref.~\cite{Contino:2006nn}.  Our assumption is that the composite sector corresponds to a strongly interacting theory, but still perturbative.  This is translated into assuming relevant NP couplings and Yukawas $\gc,\ Y_* \sim 1 - 4$.  Since we are interested in new neutral bosons and their couplings to the SM fermions, we should get the mass states for all the particles. To this end, we diagonalize analytically the Lagrangian before EWSB, and then diagonalize it again after including the EWSB effects. Since the neutral boson mixing matrix is $5 \times 5$ we choose to perform a decomposition in terms of powers of $v$. This should be taken cautiously to maintain consistency for given scale and parameters.  The new diagonalized fermions are the SM quarks and leptons and their corresponding physical composite partners, which we do not consider here. The new diagonalized bosons correspond to the photon $A$, the $Z$, $W^+$, $W^-$ and the new heavy resonances which are the neutral $Z_1$, $Z_2$ and $Z_3$ (combination of $\tilde B$, $B^*$ and $W^{3*}$) and the new charged $\tilde W,\ W^*$.  The boson fields from the old basis can be expressed as a function of the mass eigenstates as in Eq.~\ref{mixingeq} in Appendix \ref{app2}.

All new $Z_i$'s could be in principle easily tested through \zorro once the corresponding couplings are explicitly written.  We work out as an example the phenomenology of the $Z_1$ boson. Observe that $Z_1$ equals $W_3^*$ at zeroth order in the {\it vev}.  We therefore compute the required $Z_1$ couplings to the SM fermions by substituting the change of basis in the Lagrangian and extracting the couplings at second order in $v$.  Couplings of the $Z_1$ to SM fermions can be found in Eqs.~\ref{couplinglightquarks} - \ref{couplingleptons} and its decay widths to bosons in Eq.~\ref{widtheq}, in Appendix \ref{app2}. 

Once the Lagrangian is written in the mass eigenstate basis, the phenomenology depends in principle on the couplings and Yukawas ($g_*, Y_*$), the Higgs potential parameters ($\lambda_H, \mu _H$), the new sector mass parameters ($m_*, \tilde m_*, M_*$) and the mixing angles ($\theta, \varphi, \tilde \varphi$).   In the following we assume a composite fermion scale $m = \tilde m = 2$ TeV, and we vary other parameters to reproduce physical parameters compatible with experimental observations.  The SM couplings should satisfy
\begin{eqnarray}
g &=& \gc \sin (\theta)\\
Y_{SM} &=& \sin (\varphi _{\psi _L}) Y_* \sin (\varphi_{\psi _R}).
\label{yukawaestrella}
\end{eqnarray}
The SM fermion masses, which are function of the mixing angles, $Y_*$, $m_*$ and $\tilde m_*$, should reproduce the physical values.  For the given setup and interest, we scan on the parameters to fit first and third generation and define the allowed regions in parameter space. A further tuning should be performed if one would like the model to also reproduce second generation and flavor physics constraints\footnote{For NP masses in the order of a few TeV, this could be achieved for instance in the lines of Ref.~\cite{Barbieri:2012tu}.}. However, for the purposes of this work, where we aim to show the utility of \zorro and not to explore the full details of the model, the outcome of this diagonalization and mixing angles is enough to capture the essence of a $Z'$ with varying couplings to the different generations due to mass mixing terms.

For light quarks, the scan on parameter space yields similar results for up and down quarks, selecting the region of tiny mixing  angles. This is no surprising since the Yukawas and composite couplings were chosen to give strong coupling of the composite sector to the top and weak to the light fermions. For simplicity, and within the scope of our framework we can approximate $\varphi _{\psi _L} =  \varphi _{\psi _R} =0$ for all light fermions without major modifications.  Therefore, the $Z_1$ boson would couple to these light particles through its mixing with the SM bosons. For bottom and top the situation is different. The region allowed for the bottom is similar to that of the light fermions for $ \varphi _{\psi _R}$, but it is a different case for $\varphi _{\psi _L}$ which should be large since otherwise the top mass cannot be large. 
The top quark needs both $\varphi$ to be large in order to reproduce its mass. Therefore we find that $\sin( \varphi _{b_R}) \sim 0$, $\sin (\varphi_{t_R}) \sim 1$ and $\sin (\varphi _{\psi_L}) \sim 1$, which represents a large left third generation and right top coupling to the composite sector and small right bottom coupling. This yields $b_L$, $t_L$ and $t_R$ much more coupled to the composite fermions while the light SM fermions and $b_R$ are decoupled at this level of approximation.  In light of these approximations, we scan on $0.4 < \varphi_{\psi_L} < 0.9$.  Once $\varphi _{\psi_L}$ is chosen, we determine a $\varphi _{t_R}$ that reproduces the top mass within 20\% of its value.  Observe that within our approximation $Z_1$ couplings do not depend on $\varphi _{t_R}$.

Finally, we can set limits to $\theta$'s angles. Since we assume $2 < \gc < 4$, using $g_2 \approx 0.64$ and $g_Y = \sqrt{\frac{3}{5}} g_1 \approx 0.34$ and the relation $\tan \theta = \gel/\gc$ we obtain
\begin{eqnarray}
0.15 < &\theta  _2 &  < 0.3, \\
0.065 < &\theta _1 & < 0.13.
\label{titaranges}
\end{eqnarray}
In the following we extend $\theta_2$ range beyond the above  to explore its effects on the $Z_1$ phenomenology through \zorrobis.

So far we have collected the required elements for the running of \zorro to calculate the strengths ${\cal S}$ to $jj$, $bb$, $tt$, $ee$, $\mu \mu$, $\tau \tau$, $\nu \nu$, $W^+ W^-$ and $hZ$.  We fix the parameters $v=0.246$ TeV, $m_t = 0.174$ TeV, and run $M_*$ from $1-5$ TeV, $\varphi_{\psi_L}$ from $0.4-0.9$ and $\theta_2$ from $1/16-1$ (see Appendix \ref{app2}) to have a broad view in parameter space. Since we are working with $Z_1$, whose couplings have no contributions from the hypercharge sector given our approximations, parameter $\theta _1$ does not play any role (see Eqs.~\ref{couplinglightquarks} -~\ref{couplingleptons}).

\begin{figure}[!h]
\begin{center}
\includegraphics[width=0.5\textwidth]{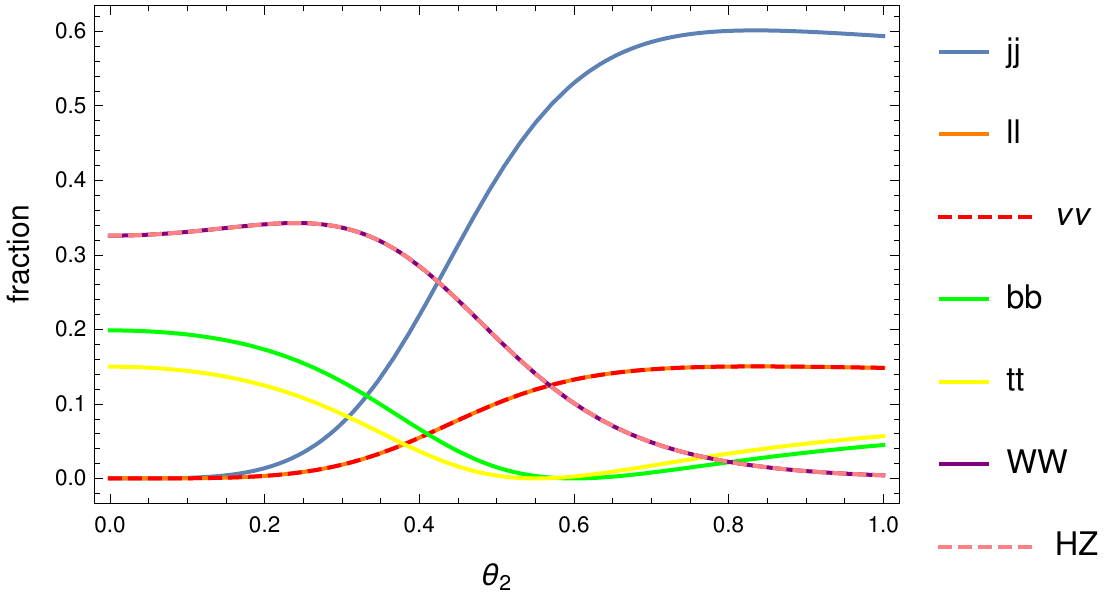}
\end{center}
	\caption{$Z_1$ Branching ratios as function of $\theta _2$ for $\varphi _{\psi _L} = 0.6$. We plotted a $\theta_2$ range wider than in Eq.~\ref{titaranges} just for phenomenology interest.}
\label{BR}
\end{figure}

Ref.~\cite{Contino:2006nn} uses $Z_1$ couplings of fermions and bosons before the last diagonalization, as a simplified approximation. We compare the branching ratios given in that work with the ones obtained above performing the full diagonalization.   We found that the widths depends on the same variables than the ones computed in Ref.~\cite{Contino:2006nn}. Being the $b$ quark massless within our approximation, we need to go to the next order in $v$ to see a dependence on $\varphi _{\psi _R}$ and $\theta _1$.  It can be seen that decays to light fermions and decays to bosons are opposite in behavior. For small $\theta _2$, the decay to light fermions is small and increases with $\theta _2$, while decay to bosons goes exactly the opposite way. 
 The heavy quarks width is small while $\varphi _{\psi _L}$ is small and increases for small $\theta _2$ when $\varphi _{\psi _L}$ is larger, assimilating to the width of the bosons. This is because if $\varphi _{\psi _L}$ is small the mixing between heavy quarks and composite heavy quarks is negligible as in the light quarks case.  Whereas for larger coupling it mimics the boson behavior, which are strongly coupled to $W^{*3}$.  In Fig.~\ref{BR} we can see the branching fractions for our calculation. As a validity check, we verify that we recover the branching fractions in Ref.~\cite{Contino:2006nn} in the corresponding limit.

\begin{figure}[h]
\begin{center}
\includegraphics[width=0.75\textwidth]{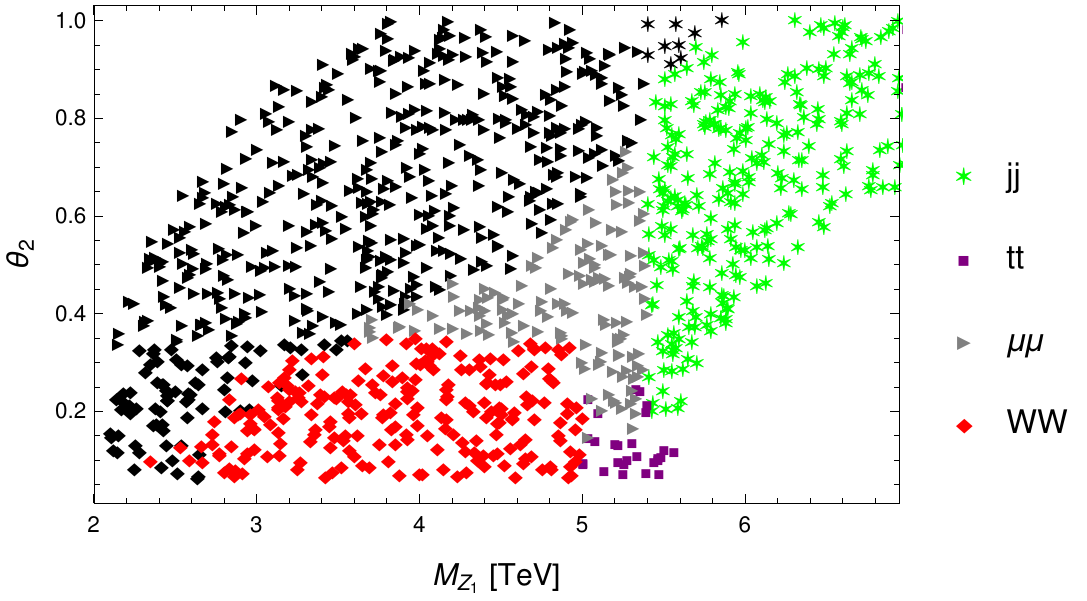}
\end{center}
\caption{Exclusion plots for the different channels as function of $\theta _2$ and $M_{Z_1}$. Color and shapes show the most sensitive channel; black meaning that the point in parameter space is excluded by the channel corresponding to its shape.}
\label{rosatitamasa}
\end{figure}

Once the above calculations, details and validity checks for the Warped/Composite model have been performed, we can easily process the model through \zorro and analyze the strength ${\cal S}$ in each channel.  We show the results for \zorro output in Fig.~\ref{rosatitamasa}.  Observe that the allowed values for the physical $Z_1$ mass obeys Eq.~\ref{zpmass} in Appendix \ref{app2}.  As it can be seen, four channels are the most sensitive depending on the region in parameter space. For $0< \theta_2 \lesssim 0.3$, it is the $WW$ channel, which agrees with the plot of the branching ratios in Fig.~\ref{BR}, since this final state has a considerable larger fraction than other channels for small $\theta _2$. The region for large mass and small $\theta_2$ in which $t\bar t$ is the most sensitive channel is because there is no data for $WW$ at these masses in the \zorro experimental cards.  In addition, for small $\theta$ the $t \bar t$ channel is preferred over $\mu\mu$, due to the BR dependence on $\theta_2$ as can be seen in Fig.~\ref{BR}. The undefined region for small $M_{Z_1}$ and $\theta_2$ is because of variations in the non-plotted $\varphi_L$ angle which modifies the branching ratios.  This is for the $\theta _2$ range indicated in Eq.~\ref{titaranges}, however, for the sake of the phenomenological interest, we also analyze the whole plot for $\theta _2>0.3$.  As $\theta_2$ increases above $0.3$ there is an interplay between the $\mu\mu$ branching ratio increase in Fig.~\ref{BR} and its sensitivity, becoming this final state the most sensitive channel (largest strength).  In some part of the parameter space --smaller masses, black color in Fig.~\ref{rosatitamasa}-- the model is discarded by $\mu\mu$, whereas in the other part is still allowed and also $\mu\mu$ is the most sensitive channel.  As $M_{Z_1}$ goes above 5.5 TeV, there are no $\mu\mu$ results and channel $jj$ becomes the most sensitive one.  The dijet final state also has a considerable increase in its branching ratio in this region. 

As it can be seen in this last result (Fig.~\ref{rosatitamasa}) for a simplified warped/composite model, the software \zorro condenses a large amount of information regarding the model itself and the experimental techniques for each specific channel, into  one number --the strength-- which yields a plot with the relevant information for the phenomenological analysis of the model.

\section{Outlook}
\label{section:5}

Along the previous paragraphs we have presented a software for performing simplified testings of BSM $Z'$ models in light of recent LHC experimental results.  We discuss in this section limitations and potential improvements of this software.

A direct limitation of \zorro is that in the aim of being general, its predictions are at leading order tree level.  This limitation cannot be avoided in a program which is for general BSM $Z'$ models, since any NLO effect is model dependent.  In fact, NLO effects through loop-level corrections depends on the full particle content in the model. However, it should be noted that if users need to include NLO effects while running \zorrobis, they can do it by computing the NLO branching ratios and adapting the \zorro input cards.  The couplings corresponding to each point in parameter space should be re-scaled accordingly, and analogously the widths.

On the other hand, there are interesting add-ons which would enhance \zorro utility.  One possibility is to schematically include low energy observables as bounds on the $Z'$ models.  Although this is a vast enterprise, it would be very appealing to have all these limits collected in an easy-to-run software such as \zorrobis.  Another compelling possibility is to include Dark Matter searches with $Z'$ as mediator, or $Z'$ as DM candidate \cite{victor}.

An important  add-on for \zorro would be to include cross-correlations in the analysis.  Since it could be the case that, because of small cross-sections, a NP model is recognized by slight excesses in different specific channels.  This undertaking would also require to cross-correlate the backgrounds since in many cases different channels have the same background.  We consider that this is a very attractive add-on for a $Z'$ finder software, and it would have to be done to fully explore the available experimental data in contrast to $Z'$ NP models.

Our code is open-source \cite{github} and any of these updates can be done by interested users.  We are willing to contribute or assist.

\section{Conclusions}
\label{section:6}

We have designed a new software {\it \zorrobis} to easily test NP models with a $Z'$ boson against experimental constraints in all $Z'$ decay channels that can be measured at LHC.  The main idea of this tool is to extract the bounds in production times branching ratio times acceptance from the ATLAS and CMS results and apply them systematically to the NP $Z'$ model.  The basic input for \zorro is a point in parameter space with the relevant phenomenological information, and the program returns as output a positive number for each channel, which is its corresponding strength.  If this number is larger than 1, means that the point in parameter space is rejected by the corresponding channel.  Moreover, the sensitivity to the given point in parameter space in each channel may be compared by studying the strength in each channel: the larger the strength, the more sensitive is the channel to reject (or observe) the NP model.   

The program does not need the user to input the model, nor run any Monte Carlo simulation.  Instead, just needs the $Z'$ mass, couplings to charged fermions and partial widths to invisible, $WW$ and $Zh$, in a simple text file, where each row is a different point in parameter space to be tested.  \zorro is intended to be as simple as possible, and therefore very fast: in a normal computer it can process 1000 points in a few seconds.   The whole program is open source written in {\tt c++} and can be used, modified and improved by other users through its {\tt Github} interface \cite{github}.

We have validated \zorro by applying it to the Sequential Standard Model (SSM) and re-obtained available results shown in Fig.~\ref{ssmplot}.  We have also worked out two other examples to explicitly show the functioning and scope of the program.   We have applied \zorro to a B-L model (Fig.~\ref{blplots}), and to a 2-sector or Warped/Composite model.  In both cases we have sketched the required work before running the software, which consists in identifying the mass eigenstate $Z'$, its couplings to SM fermions, and its width to SM bosons.  The running of \zorro on these NP models and the corresponding analysis of its output, has shown how \zorro condenses the theoretical side of the model, and the experimental techniques for each channel including its relevant backgrounds, exhibiting the power of the presented tool.  Many of these features are accordingly discussed along the text.

We have discussed benefits and issues of such a straightforward software and how in some cases the problems may have a workaround.  We have also discussed different possible improvements to \zorrobis.  We consider that including correlation between channels, low-energy physics results and Dark Matter bounds, would be very compelling.  

The main objective of the presented software \zorro is to provide a useful tool for theorist and experimentalist to test $Z'$ NP models against LHC data.

\section*{Acknowledgments}
We thank Leandro Da Rold and J. Lorenzo Díaz-Cruz for useful conversations. E.A.~thanks Manuel Szewc for testing the software, M.E.~thanks Nicol\'as Echebarrena for assistance with the code, and R.M.S.S.~thanks Diego Mouri\~no for helping with Github.

\appendix

\section{Strength example in Higgs Boson discovery}
\label{higgs}
As a concrete example on the utility of the strength $\mathcal S$ defined in Eq.~\ref{s}, we use it to study relevant features in the prediction of the 2012 Higgs boson discovery using experimental data available in 2011.  

\begin{figure}[ht]
\begin{minipage}[b]{0.48\linewidth}
\centering
\includegraphics[width=0.95\textwidth]{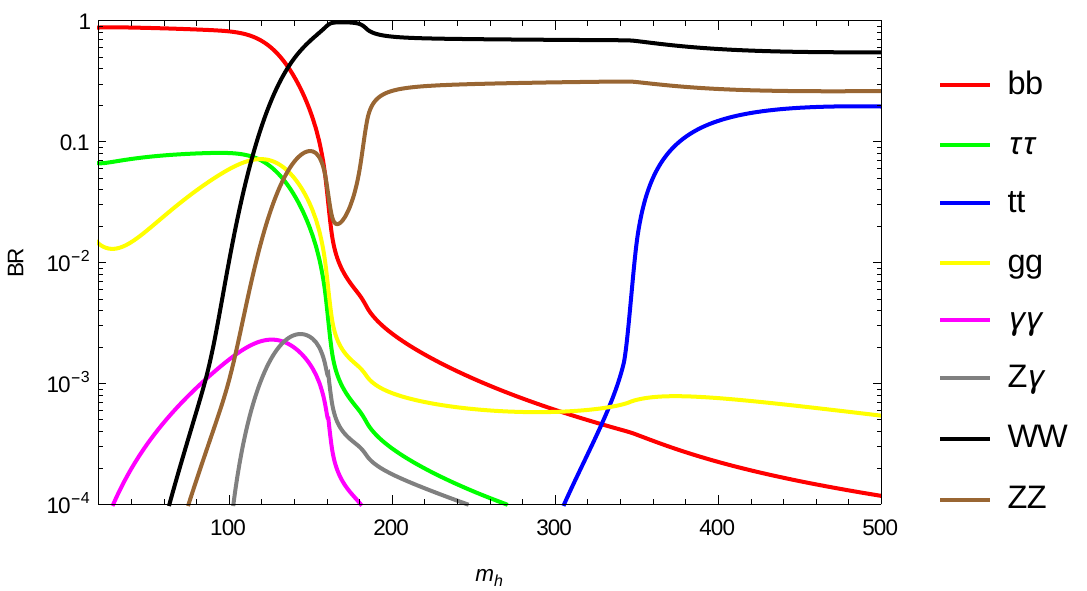}\newline
	(a)
\end{minipage}
\hspace{0.3cm}
\begin{minipage}[b]{0.48\linewidth}
\centering
\includegraphics[height=0.57\textwidth]{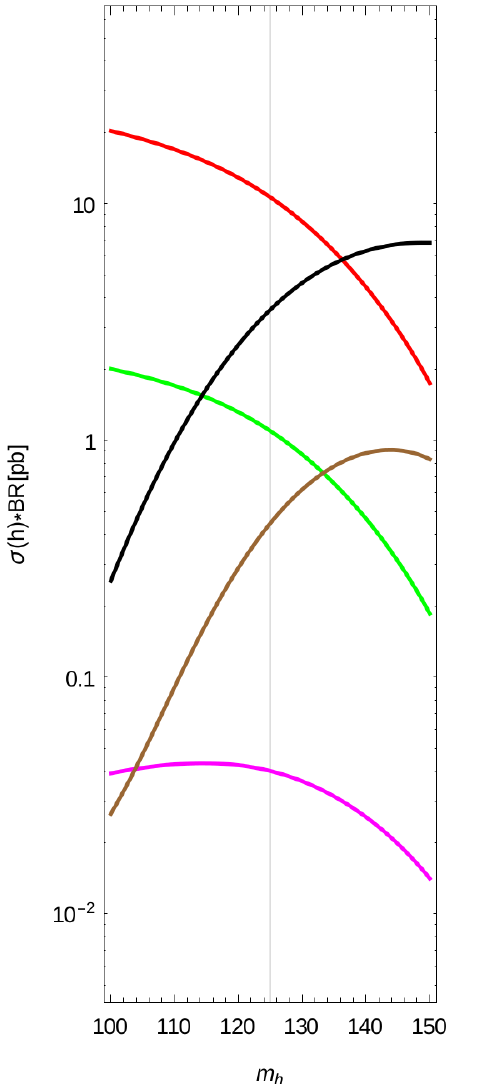}
\hskip .5cm
\includegraphics[height=0.57\textwidth]{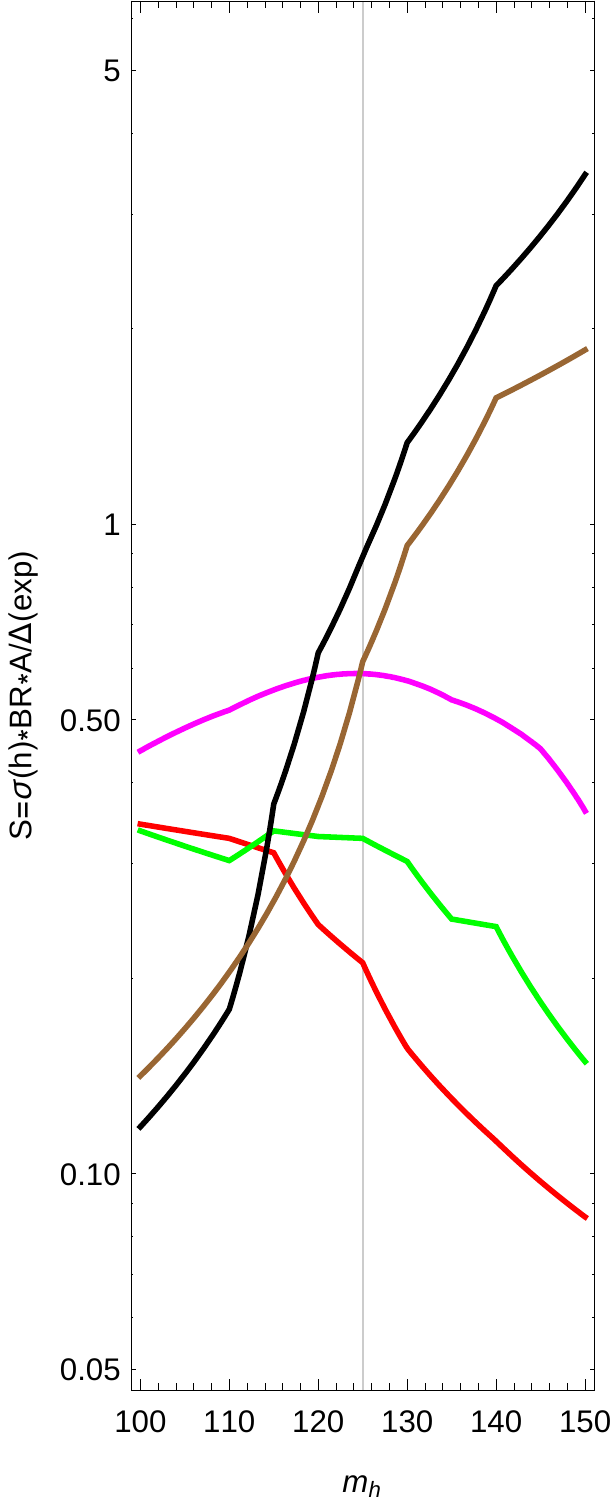} \newline
	(b)
\end{minipage}
	\caption{(a) Higgs branching ratios to different channels as function of mass.  (b) Left: Zoom to relevant region for plot in (a).  Right: Each branching ratio is divided by the sensitivity of each channel to obtain the strength $\mathcal S$.  All sensitivities and plots are from 2011, that is the previous year to Higgs discovery.  From this plot one can predict that for $m_h=125$ GeV, $WW$ is the most sensitive channel to measure the Higgs boson, and $\gamma\gamma$ and $ZZ$ are equally sensitive.  See text for explanations.}
	\label{brs}
\end{figure}

We show in Fig.~\ref{brs}a the predicted branching ratios of the Higgs boson as a function of its (then unknown) mass.  Although $b\bar b$ is the main decay channel for a 125 GeV Higgs boson, its background is very large at the LHC, and therefore it is an unlikely channel for its discovery.  On the other hand $WW$ and $ZZ$ have a non negligible branching ratio and background not too large, albeit that the $WW$ channel does not have the power to predict the resonance mass due to the missing energy leaving the detector as neutrinos.  The particular case is $\gamma\gamma$ channel which has a tiny branching ratio, but also a tiny background, therefore being a very likely channel for discovery.  This qualitative discussion can be taken to a quantitative analysis by simply using the strength $\mathcal S$. 

In order to perform a quantitative analysis using $\mathcal S$, we extract a zoom of the Higgs decay channels for the relevant region in masses and channels in Fig.~\ref{brs}b left panel.  To these results we apply the computation of the strength.  We multiply by the acceptance and divide by the 2011 LHC  predicted 95\% C.L.~ limits in cross-sections $\times$ branching ratio $\times$ acceptance in each channel.  We obtain these latter limits by either the numeric available data or by digitalizing the 2011 available experimental results in $b\bar b$ \cite{CMS:2011tpa} (7 TeV, 1.1 fb$^{-1}$) , $ZZ$ (7 TeV, 4.7 fb$^{-1}$) \cite{CMS:2011qta}, $WW$ \cite{CMS:2011cta} (7 TeV, 4.6 fb$^{-1}$), $\tau\tau$ \cite{CMS:2011kta} (7 TeV, 4.6 fb$^{-1}$) and $\gamma\gamma$ \cite{CMS:2011nqa} (7 TeV, 1.09 fb$^{-1}$), where all data has been normalized to luminosity 4.7 fb$^{-1}$ using Eq.~\ref{luminosity}.  The outcome of the strength is plotted in Fig.~\ref{brs}b right panel.  In this figure it can be seen which is the 2011 discovery prediction using strength: the first channel where a 125 GeV is going to be observed is $WW$, and then shortly after in $\gamma\gamma$ and $ZZ$ simultaneously. 

\begin{figure}[h!]
        \begin{center}
\includegraphics[width=.8\textwidth]{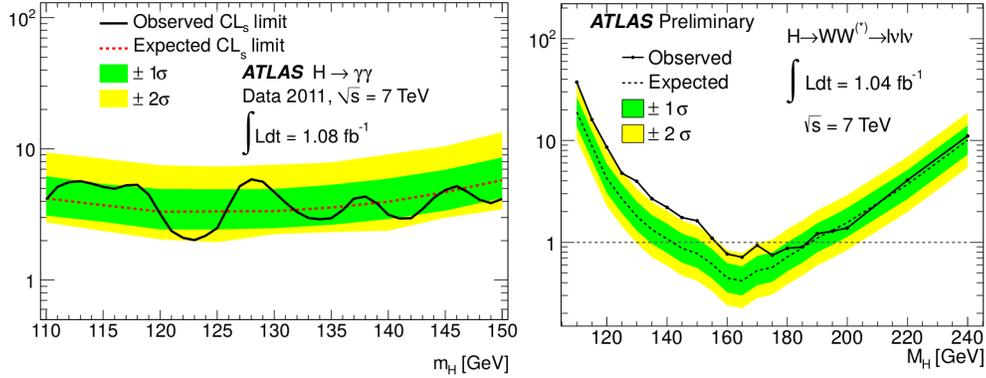}
        \end{center}
	\caption{Comparison of limits in Higgs boson discovery for $\gamma\gamma$ \cite{Aad:2011ww} and $WW$ \cite{ATLAS:2011hia} channels at same energy and luminosity conditions in year 2011.  As predicted by the strength $\mathcal S$ analysis, the $WW$ channel is more sensitive than $\gamma\gamma$, although it does not have a resonant peak signal because of the missing energy taken away by neutrinos.}
        \label{2011}
\end{figure}

The Higgs boson was discovered in 2012 using 4.8 fb$^{-1}$ (7 TeV) and 5.8 fb$^{-1}$ (8 TeV) of luminosity by ATLAS \cite{Aad:2012tfa} and 5.1 fb$^{-1}$ (7 TeV) and 5.3 fb$^{-1}$ (8 TeV) by CMS \cite{Chatrchyan:2012xdj}.  Although the more impressive channels were $\gamma\gamma$ and $ZZ$ because a resonant peak could be distinguished at 125 GeV, the excess in $WW$ was available before, as predicted by the strength $\mathcal S$ analysis in previous paragraph and in Fig~\ref{brs}b.  This can be seen in Fig.~\ref{2011}, where we have put aside for comparison the observed and expected limits for $\gamma\gamma$ and $WW$ at same luminosity and energy conditions.  The $WW$ channel is already measuring the Higgs excess, although it does not show any resonant peak because of the missing energy in the decay products.

Along this concrete example of the Higgs boson discovery we have shown the usage and utility of the strength.  We present the agreement between the 2011 predictions and the 2012 experimental results, as an estimate example on how could work a simple analysis using strength $\mathcal S$ in predicting features of an eventually yet unobserved new $Z'$ resonance using \zorro.  

\section{$Z_1$ couplings and widths to SM particles}
\label{app2}
In this Appendix we expand some of the details of the $Z_1$ neutral gauge boson in the Warped/Composite framework. 

We perform the expansion of the fermions eigenvectors to second order in $v/M_*$. In the case of the bosons, the first order is zero because in the original matrix the {\it vev} already appears quadratic, and hence we also keep here to second order. The boson fields from the previous basis, written as a function of the new mass eigenstates read, up to a normalization constant,
{\small \begin{equation}
	\begin{aligned}
W_3 &= \frac{g_Y}{\sqrt{g_2 ^2 + g_Y^2}} A + \frac{g_2}{\sqrt{g_2 ^2 + g_Y^2}} Z, \\
B &= \frac{g_2}{\sqrt{g_2 ^2 + g_Y^2}} A - \frac{g_Y}{\sqrt{g_2 ^2 + g_Y^2}} Z - \frac{c_2 ^3 g_2 (g_2 ^2 + g_Y^2)v^2 }{4 M_* ^2 s_2 \sqrt{g_2 ^2 + g_Y^2}} Z_1   \\
&+ \frac{c_1 ^3 g_Y (g_2 ^2 + g_Y^2)v^2 }{4 M_* ^2 s_1 \sqrt{g_2 ^2 + g_Y^2}} Z_2 + \frac{g_Y (g_2 ^2 + g_Y^2)v^2 }{2 \sqrt{6} M_* ^2 s_1 \sqrt{g_2 ^2 + g_Y^2}} Z_3,  \\
\tilde B &= \frac{g_2 g_Y v^2 }{2 \sqrt{6} M_*^2 s_1} A - \frac{g_Y^2 v^2}{2 \sqrt{6} M_*^2 s_1} Z - \frac{c_2^3 g_2 g_Y v^2}{2 \sqrt{6} M_* ^2 s_1 s_2 (1 - c_2^2  ) } Z_1 + \frac{c_1^3 g_Y^2 v ^2}{2 \sqrt{6} M_* ^2 s_1 (1 - c_2^2  ) } Z_2 -Z_3, \\
B^* &= \frac{c_1^3 g_2 g_Y v^2}{4 M_*^2 s_1} A - \frac{c_1^3 g_Y^2 v^2}{4 M_*^2 s_1} Z - \frac{c_1^3 c_2^3 g_2 g_Y v^2}{4 M_*^2 s_1 s_2 (c_1^2 - c_2^2)} Z_1 - Z_2 + \frac{c_1^3 g_Y^2 v^2}{2 \sqrt{6} M_*^2 s_1^2 (c_1^2 - 1)} Z_3,  \\
W_3^* &= -\frac{c_2^3 g_2^2 v^2}{4 M_*^2 s_2} A + \frac{c_2^3 g_2 g_Y v^2}{4 M_*^2 s_2} Z - Z_1 + \frac{c_1^3 c_2^3 g_2 g_Y v^2}{4 M_*^2 s_1 s_2 (c_1^2 - c_2^2) } Z_2 + \frac{c_1^3 g_2 g_Y v^2}{2 \sqrt{6} M_*^2 s_1 s_2 (1 - c_2^2)} Z_3,
	\end{aligned}
	\label{mixingeq}
\end{equation}}
where the parameters are the same as those used in Ref.~\cite{Contino:2006nn}. The physical masses of the $Z$ and the photon are the usual while for the new $Z' _i$ they are given by
\begin{eqnarray}
	\begin{aligned}
m_{Z_1} ^2 &=& \frac{M_*^2}{c_2 ^2} + \frac{c_2^2 g_2^2 v^2}{4 s_2 ^2},  \\
m_{Z_2} ^2 &=& \frac{M_*^2}{c_1 ^2} + \frac{c_1^2 g_Y^2 v^2}{4 s_1 ^2},  \\
m_{Z_3} ^2 &=& M_*^2 + \frac{g_Y^2 v^2}{6 s_1 ^2}. 
	\end{aligned}
\label{zpmass}
\end{eqnarray}
We can obtain the $Z_1$ couplings to bosons and fermions by substituting the old bosons for the mass states after the diagonalization. We extract the couplings at second order in $v/M_*$ and work in the region $\theta > (v/M_*)^2$ to assure the validity of the perturbative expansion.  Since in our scanning 1 TeV is the lowest bound for the $M_*$ parameter, we take $\theta_2 > 1 / 16 $. 

Since we consider all fermions but the top as massless, we use $Y_f = 0$ except $Y_t = 1$. Analogously, we take $\varphi _{b_R} = 0$. This simplifies the model and yields a better understanding of the couplings behavior.  We can see that all light fermions have a similar coupling, with the Left ones having an extra term coming from their $SU(2)$ interaction. The top quark has extra terms in both chiralities because of the non vanishing Yukawa, and the Left chirality because of $SU(2)$.  The same goes for the bottom, except that since this is a massless quark, its Yukawa being zero makes zero extra terms and also $\varphi _{bR} = 0$. 

Couplings to light quarks, are given by
\begin{equation}
	\begin{aligned}
g_{u_L} &= - \frac{g_2 g_y (g_2^2 + g_y^2) \cos (\theta _2) ^2 \cot (\theta _2) v^2 }{24 \sqrt{g_y^2 + g_2^2} M_*^2} + \frac{1}{2} g_2 \tan (\theta _2), \\
g_{u_R} &= - \frac{g_2 g_y (g_2^2 + g_y^2) \cos (\theta _2) ^2 \cot (\theta _2) v^2 }{6 \sqrt{g_y^2 + g_2^2} M_*^2}, \\
g_{d_L} &= - \frac{g_2 g_y (g_2^2 + g_y^2) \cos (\theta _2) ^2 \cot (\theta _2) v^2 }{24 \sqrt{g_y^2 + g_2^2} M_*^2} - \frac{1}{2} g_2 \tan (\theta _2), \\
g_{d_R} &= \frac{g_2 g_y (g_2^2 + g_y^2) \cos (\theta _2) ^2 \cot (\theta _2) v^2 }{12 \sqrt{g_y^2 + g_2^2} M_*^2}.
	\end{aligned}
\label{couplinglightquarks}
\end{equation}
Within our level of approximation we consider the above couplings for first and second generations.

The couplings to the third generation of quarks are given in Eq.~\ref{couplingheavyquarks}, where the top is considered heavy and the bottom massless.
\begin{equation}
	\begin{aligned}
		g_{t_L} =& - \frac{g_2 g_y (g_2^2 + g_y^2) \cos (\theta _2) ^2 \cot (\theta _2) v^2 }{24 \sqrt{g_y^2 + g_2^2} M_*^2} \\
		&+ \frac{v g_2 Y_t \cot (\varphi _{\psi _L})}{2 \sqrt{2} m_*} \cos (\varphi _{\psi _L}) \csc (\theta _2) \sec (\theta _2) \sin (\varphi _{\psi _L})  \\
		&- \frac{1}{2} g_2 \left( \cot (\theta _2) \sin (\varphi _{\psi _L})^2 - \tan (\theta _2) \cos (\varphi _{\psi _L}) ^2 \right) \\
		&- \frac{v^2 g_2 Y_t^2 \cot (\varphi _{\psi _L}) ^2 }{4 m_* ^2} \left( \cot (\theta _2) \cos (\varphi _{\psi _L}) ^2 - \tan (\theta _2) \sin (\varphi _{\psi _L})^2 \right), \\
		g_{t_R} =& - \frac{g_2 g_y (g_2^2 + g_y^2) \cos (\theta _2) ^2 \cot (\theta _2) v^2 }{6 \sqrt{g_y^2 + g_2^2} M_*^2}, \\
		g_{b_L} =& - \frac{g_2 g_y (g_2^2 + g_y^2) \cos (\theta _2) ^2 \cot (\theta _2) v^2 }{24 \sqrt{g_y^2 + g_2^2} M_*^2} + \frac{1}{2} g_2 \left( \cot (\theta _2) \sin (\varphi _{\psi _L})^2 - \tan (\theta _2) \cos (\varphi _{\psi _L}) ^2 \right), \\
		g_{b_R} =& \frac{g_2 g_y (g_2^2 + g_y^2) \cos (\theta _2) ^2 \cot (\theta _2) v^2 }{12 \sqrt{g_y^2 + g_2^2} M_*^2}.
	\end{aligned}
\label{couplingheavyquarks}
\end{equation}
Here $m_*=\frac{m}{\cos (\varphi _{\psi _L}) }$  is the diagonalized mass.

Finally, couplings to all leptons are given in Eq.~\ref{couplingleptons}, where the three generations are considered light.
\begin{equation}
	\begin{aligned}
		g_{e_L} =& \frac{g_2 g_y (g_2^2 + g_y^2) \cos (\theta _2) ^2 \cot (\theta _2) v^2 }{8 \sqrt{g_y^2 + g_2^2} M_*^2} - \frac{1}{2} g_2 \tan (\theta _2),\\
g_{e_R} =& \frac{g_2 g_y (g_2^2 + g_y^2) \cos (\theta _2) ^2 \cot (\theta _2) v^2 }{4 \sqrt{g_y^2 + g_2^2} M_*^2}, \\
g_{\nu_L} =& \frac{g_2 g_y (g_2^2 + g_y^2) \cos (\theta _2) ^2 \cot (\theta _2) v^2 }{8 \sqrt{g_y^2 + g_2^2} M_*^2} - \frac{1}{2} g_2 \tan (\theta _2), \\
g_{\nu_R} =& 0. 
	\end{aligned}
\label{couplingleptons}
\end{equation}


%


In order to use the \zorro software in this framework, we need to find the decay width of the $Z_1$ to SM bosons. The easiest way is considering the Equivalence Theorem \cite{Riesselmann:1995gv}, which states that the longitudinal $W$ and $Z$ are equivalent to the corresponding Goldstone Bosons in the high energy limit. As a consequence, the widths are given by
\begin{eqnarray}
\Gamma (W_3^* \rightarrow Zh) = \Gamma (W_3^* \rightarrow W^+ W^-) = \frac{g_2^2 M_*}{192 \pi} \cot (\theta _2) ^2.
\end{eqnarray}
Notice that these widths corresponds to the $W_3^*$ decay, and not the new mass states. To get the width corresponding to the physical states, one needs to perform the diagonalization and extract the couplings of the new $Z_1$ to the SM bosons. This is not straightforward, since one should also consider the gauge Lagrangian and the rotations of the charged bosons. For the sake of simplicity, we use the approximation for high masses in which $W^{*}_3$ is mainly composed of $Z_1$ (see Eq.~\ref{mixingeq}), while the contribution from $Z_2$ and $Z_3$ quickly decays to zero. A larger $M_*$ assures that $v/M_*$ is small and therefore a fast convergence. For the case of smaller $Z'$ masses, we found that for small angles the three neutral boson masses are of the same order.  We therefore choose to analyze $Z_1$ phenomenology and approximate the following decay widths as
\begin{eqnarray}
\Gamma (Z_1 \rightarrow Zh) = \Gamma (Z_1 \rightarrow W^+ W^-) = \frac{g_2^2 M_{Z_1}}{192 \pi} \cot (\theta _2) ^2.
\label{widtheq}
\end{eqnarray}



\bibliographystyle{JHEP}
\bibliography{biblio}

\end{document}